\documentclass[a4,numbers,sort&compress]{elsarticle}%
\usepackage{amsfonts}
\usepackage{amssymb,amsmath,amsthm,graphics,color}
\usepackage{graphics}
\usepackage{graphicx}
\usepackage{epsfig}
\usepackage{fancybox}
\usepackage{caption}
\usepackage{tikz,mathpazo}
\usepackage{multirow,booktabs}
\usepackage{amsmath}
\usepackage{amssymb}
\usepackage{geometry}
\usepackage[pagewise]{lineno}%
\usetikzlibrary{shapes.geometric, arrows}
\usetikzlibrary{calc}
\journal{}

\newtheorem{theorem}{Theorem}[section]
\newtheorem{corollary}{Corollary}[section]
\newtheorem{remark}{Remark}[section]
\newtheorem{example}{Example}[section]
\newtheorem{lemma}{Lemma}[section]
\begin{document}
\begin{frontmatter}
\title{A novel analysis approach of uniform persistence for a COVID-19 model with quarantine and standard incidence rate}
% or include affiliations in footnotes:
\author[mymainaddress,mysecondaryaddress]{Songbai Guo}
\ead{guosongbai@bucea.edu.cn}
\author[mymainaddress]{Yuling Xue}
\ead{xyl981274902@163.com}
\author[mysecondaryaddress1]{Xiliang Li\corref{mycorrespondingauthor}}
\ead{lixiliang@amss.ac.cn}
\author[mymainaddress0,mysecondaryaddress,mysecondaryaddress0]{Zuohuan Zheng}
\ead{zhzheng@amt.ac.cn}
\cortext[mycorrespondingauthor]{Corresponding author.}
\address[mymainaddress]{School of Science, Beijing University of Civil Engineering and Architecture, Beijing 102616, P. R. China}
\address[mymainaddress0]{School of Mathematics and Statistics, Hainan Normal University, Haikou 571158, P. R. China}
\address[mysecondaryaddress]{Academy of Mathematics and Systems Science, Chinese Academy of Sciences,
Beijing 100190, P. R. China}
\address[mysecondaryaddress1]{School of Mathematics and Information Science, Shandong Technology and Business University, Yantai 264005, P. R. China}
\address[mysecondaryaddress0]{School of Mathematical Sciences, University of Chinese Academy of Sciences,
Beijing 100049, P. R. China}
\begin{abstract}
A coronavirus disease 2019 (COVID-19) model with quarantine and standard incidence rate is first developed, then a novel analysis approach for finding the ultimate lower bound of COVID-19 infectious individuals is proposed, which means that the COVID-19 pandemic is uniformly persistent if the control reproduction number $\mathcal{R}_{c}>1$. This approach can be applied to other related biomathematical models, and some existing works can be improved by using that. In addition, the COVID-19-free equilibrium $V^0$ is locally asymptotically stable (LAS) if $\mathcal{R}_{c}<1$ and linearly stable if $\mathcal{R}_{c}=1$, respectively;
while $V^0$ is unstable if $\mathcal{R}_{c}>1$.
\end{abstract}
\end{frontmatter}

%\linenumbers

\section{Introduction}
At present, the COVID-19 caused by the severe acute respiratory syndrome
coronavirus 2 (SARS-CoV-2), which emerged in December 2019 has spread around
the globe. As of September 2, 2022, there have been cumulatively 601,189,435
confirmed cases in the world, of which 6,475,346 deaths \cite{WHO22}. The
COVID-19 not only inflicts a global public health crisis, but also has a major
impact on the normal life of humans \cite{Roda03}. In the early stages of the
COVID-19 pandemic, some large-scale activities exacerbated the spread of the
epidemic \cite{Zeller.19}. Following World Health Organization (WHO) report,
COVID-19 can be spread by contact and droplets, airborne and contaminant
transmission, among other means. Available evidence suggests that SARS-CoV-2
is passed from human-to-human mainly through respiratory droplets and contact
routes \cite{WHO20}. If domestic animals or wild animals become the host of
SARS-CoV-2, then COVID-19 will pose a greater threat to humans \cite{Otto5}.

In the process of epidemic prevention and control, mathematical modeling
methods can help us understand the interaction between different
epidemiological factors, thereby helping to control the transmission of this
epidemic \cite{Amer06}. Infected individuals are divided into symptomatic
infections, and asymptomatic infections who have a positive nucleic acid test
but do not show any symptoms \cite{Gao07}. Since asymptomatically infected
individuals do not know that they have been infected by the virus, the
transmission caused by these people accounts for the vast majority
\cite{Gumel02}. Thus, the mathematical model of COVID-19 with asymptomatic
transmission will be more reasonable. Analyzing the dynamic behavior of the
infectious disease model helps us comprehend the long-term behavior of the
mathematical model so as to more effectively control the spread of the disease
\cite{Amer06,Khan08}. Kiouach et al. \cite{Kiouach21} established a SQEAIHR
(S: susceptible individuals, Q: quarantined individuals, E: exposed
individuals, A: asymptomatically infected individuals, I: symptomatically
infected individuals, H: hospitalized individuals, R: recovered individuals)
mathematical model for COVID-19 and demonstrated that this model is uniformly
persistent if $R_{0}>1$, which means that COVID-19 will persist in the
population. Zhang et al. \cite{Zhang20} developed a stochastic model of
COVID-19 and found some sufficient conditions for the persistence or the
extinction of the disease. Cui et al. \cite{Cui20} gave a thorough analysis
for the global stability of equilibria of a hepatitis C virus model with acute
and chronic infections. Cheng et al. \cite{Cheng21} investigated the global
stability of equilibria of a SIQS (I: infected individuals) model with
quarantine measure under some conditions. Jiang et al. \cite{Jiang21} used
SEIAR and SEIA-CQFH (C: community isolation, Q: quarantine point isolation, F:
Fangcang shelter hospitals, H: designated hospitals) models to assess
qualitatively the effects of joint measures led by Fangcang shelter hospitals
in response to COVID-19 pandemic in Wuhan, China. Mohsen et al.
\cite{Mohsen21} believed that one of the reasons for the spread of COVID-19 is
immigration, thus they proposed a system that takes into account the impact of
immigration and quarantine. Their findings suggest that the disappearance of
the disease is due to the implementation of quarantine measures.

Recently, Bai et al. \cite{Bai21} established the SEIAQR model for the spread
of mumps:
\begin{equation}%
\begin{split}
\dot{S}(t)  &  =\lambda-\beta S(t)(aE(t)+I(t)+bA(t))-dS(t),\\
\dot{E}(t)  &  =\beta S(t)(aE(t)+I(t)+bA(t))-(c+d)E(t),\\
\dot{I}(t)  &  =pcE(t)-(q+r+d)I(t),\\
\dot{A}(t)  &  =(1-p)cE(t)-(r+d)A(t),\\
\dot{Q}(t)  &  =qI(t)-(r+d)Q(t),\\
\dot{R}(t)  &  =rI(t)+rA(t)+rQ(t)-dR(t),
\end{split}
\label{mumod}%
\end{equation}
and gave a complete analysis for the global stability of the disease-free
equilibrium and the unique pandemic equilibrium of model \eqref{mumod}. In
this model \eqref{mumod}, $q$ is the quarantined rate of symptomatic
infections and $r$ stands for the recovery rate, and the descriptions of all
other parameters are listed in Tab. \ref{tab1}. From the transmission
characteristics of COVID-19, the disease can be transmitted by exposed
individuals, symptomatically and asymptomatically infected individuals
\cite{Cui201,Jiang21,Wang20,Wang21}. In fact, model \eqref{mumod} is also in
compliance with the propagation mechanism of COVID-19. Following the
discussion of McCallum et al. \cite{McCallum01}, the standard incidence rate
can better reflect the transmission of a pathogen. Thus, we will develop a
COVID-19 model \eqref{mod1} (also see Fig. \ref{fig1}) with standard incidence
rate on the basis of \cite{Bai21}.

Our model differs from model \eqref{mumod} in three ways. Firstly, the
contributions of the interaction among susceptible individuals $S$, exposed
individuals $E$, symptomatically infected individuals $I$ and asymptomatically
infected individuals $A$ to the growth rate of exposed individuals are no
longer accounted for by the mass action term $S(t)(aE(t)+I(t)+bA(t))$, which
has been replaced with the standard incidence term
$S(t)(aE(t)+I(t)+bA(t))/N(t)$, where%
\[
N(t)=S(t)+E(t)+I(t)+A(t)+Q(t)+R(t).
\]
Secondly, the quarantined rate of asymptomatic infections is added and
different from that of symptomatic infections. Thirdly, the recovery rates of
symptomatic infections, asymptomatic infections and quarantine are different.
At present, numerical results of COVID-19 models with standard incidence rates
are abundant, while dynamics analysis is rare. Our purpose is to present a
more refined approach of uniform persistence of model \eqref{mod1} by using a
thorough analysis, which can give some refined estimates to the ultimate lower
bounds of solutions of the model.

The rest of this paper is structured as follows. In Section 2, the model
formulation is given. In Section 3, the control reproduction number $R_{c}$ is
calculated and the existence condition of the COVID-19 equilibrium is
obtained. In Section 4, the stability of the COVID-19-free equilibrium is
analyzed, and a complete analysis approach is proposed for the uniform
persistence of model \eqref{mod1}. Meanwhile, some explicit estimations on the
ultimate lower bound of COVID-19 individuals are acquired, and some examples
are given to illustrate our main result. Finally, a brief conclusions section
completes this paper.

\section{Model formulation}

We divide the total population $N$ into six subclasses: susceptible
individuals $S$, exposed individuals $E$, symptomatically infected individuals
$I$, asymptomatically infected individuals $A$, quarantined individuals $Q$
and recovered individuals $R$. To this end, a flow chart of COVID-19
transmission model is shown in Fig. 1, where all parameters of this model are
positive and their definitions are listed in Tab. \ref{tab1}, and $p\in(0,1).$
\begin{figure}[ptbh]
\centering
\includegraphics[width=10cm]{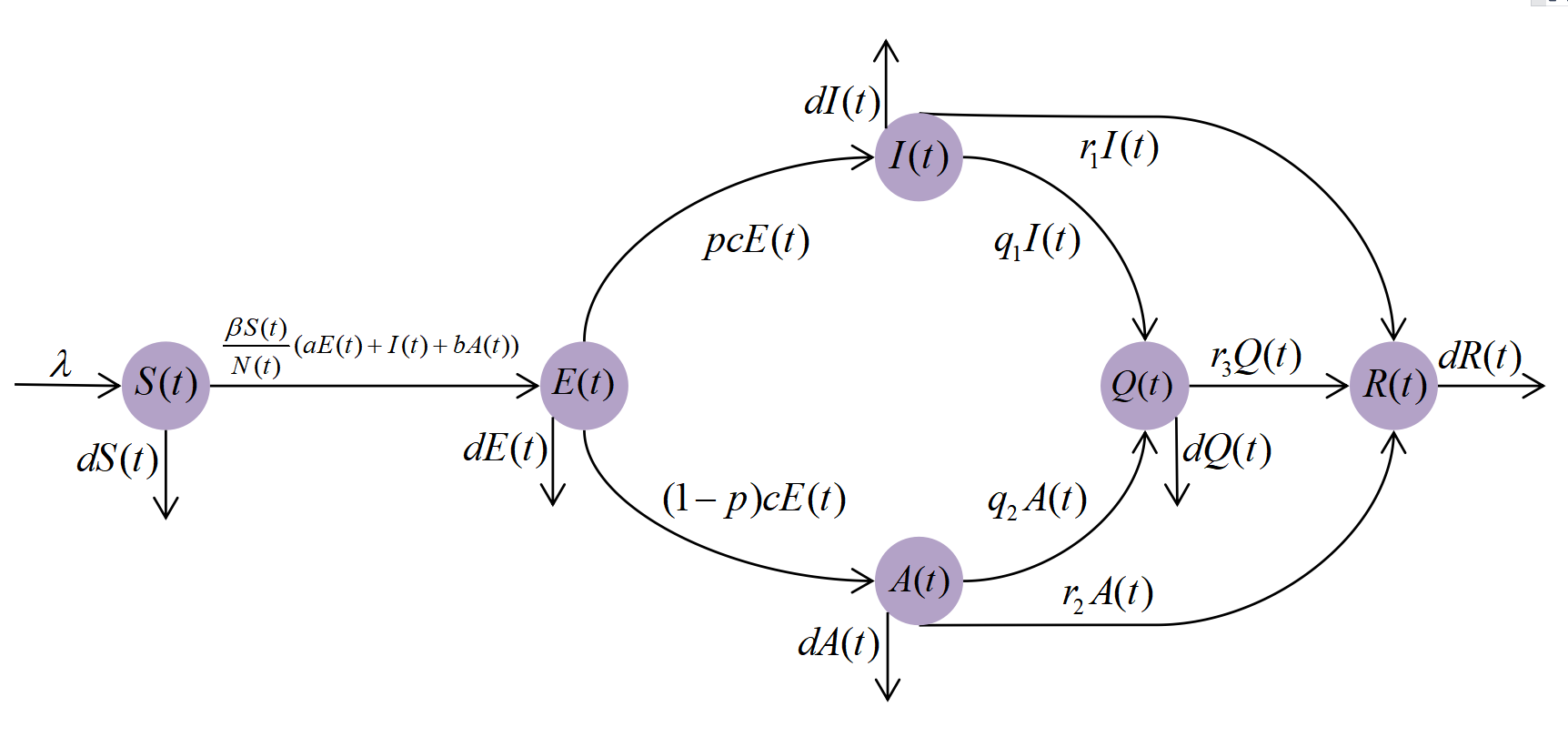} \caption{Flow chart of the COVID-19
transmission model}%
\label{fig1}%
\end{figure}

From Fig. \ref{fig1}, the COVID-19 transmation model is as follows,%
\begin{equation}%
\begin{split}
\dot{S}(t)  &  =\lambda-\beta\frac{S(t)}{N(t)}(aE(t)+I(t)+bA(t))-dS(t),\\
\dot{E}(t)  &  =\beta\frac{S(t)}{N(t)}(aE(t)+I(t)+bA(t))-(c+d)E(t),\\
\dot{I}(t)  &  =pcE(t)-(q_{1}+r_{1}+d)I(t),\\
\dot{A}(t)  &  =(1-p)cE(t)-(q_{2}+r_{2}+d)A(t),\\
\dot{Q}(t)  &  =q_{1}I(t)+q_{2}A(t)-(r_{3}+d)Q(t),\\
\dot{R}(t)  &  =r_{1}I(t)+r_{2}A(t)+r_{3}Q(t)-dR(t).
\end{split}
\label{mod1}%
\end{equation}
\begin{table}[th]
\caption{Definition of parameters in model \eqref{mod1}.}%
\label{tab1}%
\begin{tabular}
[c]{cl}%
\toprule Parameter & Definition\\
\midrule $\lambda$ & The birth rate of susceptible individuals\\
$d$ & The natural death rate\\
$\beta$ & The transmission rate of COVID-19\\
$a$ & The regulatory factor for infection probability of exposed individuals\\
$b$ & The regulatory factor for infection probability of asymptomatically
infected individuals\\
$c$ & The transfer rate of exposed individuals to other infected individuals\\
$p$ & The transition probability of symptomatically infected individuals\\
$q_{1}$ & The quarantined rate of symptomatically infected individuals\\
$q_{2}$ & The quarantined rate of asymptomatically infected individuals\\
$r_{1}$ & The recovery rate of symptomatically infected individuals\\
$r_{2}$ & The recovery rate of asymptomatically infected individuals\\
$r_{3}$ & The recovery rate of quarantined individuals\\
\bottomrule &
\end{tabular}
\end{table}

In virtue of the general theory of ordinary differential equations (see, e.g.,
\cite{Chepyzov02,Hale80}), we know that model \eqref{mod1} is well-posed and
dissipative in the set
$$
D=\left\{\phi=(\phi_{1},\phi_{2},\phi_{3},\phi_{4},\phi_{5},\phi_{6})^{T}
\in\mathbb{R}_{+}^{6}:
\sum_{i=1}^{6}\phi_{i}>0\right\}
$$
positively invariant
for the model system, where $\mathbb{R}_{+}=[0,\infty)$. Thus, we will analyze
the global dynamics of system \eqref{mod1} in $D$.

\section{Existence of pandemic equilibrium}

It is clear to see that system \eqref{mod1} always has a COVID-19-free
equilibrium $V^{0}=(S^{0},0,0,0,0,0)^{T}$, where $S^{0}=\lambda/d$. To get the
existence of the COVID-19 equilibrium $V^{\ast}=(S^{\ast},E^{\ast},I^{\ast
},A^{\ast},Q^{\ast},R^{\ast})^{T}$ of system \eqref{mod1}, we first calculate
the control reproduction number
\begin{equation}
\mathcal{R}_{c}=\frac{a\beta}{c+d}+\frac{pc\beta}{(c+d)B_{1}}+\frac
{bc\beta(1-p)}{(c+d)B_{2}}, \label{crn}%
\end{equation}
by using the method in \cite{Driessch13}, where $B_{i}:=q_{i}+r_{i}+d$,
$i=1,2.$ Here, the first term can be expressed as that an exposed individual
can averagely infect $a\beta$ susceptible individuals in a unit time, and the
average duration of the exposure period is $1/(c+d)$. And the second term can
be expressed as that the exposed individuals with $pc/(c+d)$ can be
transformed into the symptomatically infected individuals, a symptomatically
infected individual can averagely infect $\beta$ susceptible individuals in a
unit time, and the average duration of symptomatic infection is $1/B_{1}$.
While the third term can be expressed as that the exposed individuals with
$(1-p)c/(c+d)$ can be transformed into the asymptomatically infected
individuals, an asymptomatically infected individual can averagely infect
$b\beta$ susceptible individuals in a unit time, and the average duration of
asymptomatic infection is $1/B_{2}$.

\begin{lemma}
\label{lem0} System \eqref{mod1} possesses a unique COVID-19 equilibrium
$V^{\ast}$ if and only if $\mathcal{R}_{c}>1$.
\end{lemma}

\proof Let the right-hand sides of system \eqref{mod1} equal zero, it follows
$N=S^{0}.$ Thus, we have
\begin{align}
S  &  =\frac{\lambda S^{0}B_{1}B_{2}}{B_{1}B_{2}\left(  dS^{0}+\beta
aE\right)  +E\left[  B_{2}p+B_{1}b\left(  1-p\right)  \right]  \beta c}%
=\frac{\lambda-\left(  c+d\right)  E}{d},\label{eq3}\\
I  &  =\frac{pcE}{B_{1}},\text{ }A=\frac{(1-p)cE}{B_{2}},\text{ }Q=\frac
{q_{1}pcE}{B_{1}\left(  r_{3}+d\right)  }+\frac{q_{2}\left(  1-p\right)
cE}{B_{2}\left(  r_{3}+d\right)  },\nonumber\\
R  &  =\frac{r_{1}pcE}{dB_{1}}+\frac{r_{2}\left(  1-p\right)  cE}{dB_{2}%
}+\frac{r_{3}q_{1}pcE}{dB_{1}\left(  r_{3}+d\right)  }+\frac{r_{3}q_{2}\left(
1-p\right)  cE}{dB_{2}\left(  r_{3}+d\right)  }\nonumber
\end{align}
According to \eqref{eq3}, there holds%

\[
\left(  c+d\right)  B_{1}B_{2}E(a_{1}E-a_{2})=0,
\]
where
\[
a_{1}=\left(  c+d\right)  R_{c}>0,\text{ }a_{2}=\lambda\left(  R_{c}-1\right)
.
\]
Therefore, system \eqref{mod1} possesses a unique pandemic equilibrium
$V^{\ast}\gg\mathbf{0}$ if and only if $0<E^{\ast}=\frac{a_{2}}{a_{1}}%
<\frac{\lambda}{c+d},$ namely, $\mathcal{R}_{c}>1$.

\begin{remark}
\label{rem1} It is not difficult to find that $S^{0}=S^{\ast}+E^{\ast}%
+I^{\ast}+A^{\ast}+Q^{\ast}+R^{\ast}$ and $\mathcal{R}_{c}=S^{0}/S^{\ast}$ for
$\mathcal{R}_{c}>1.$
\end{remark}

\section{Stability and uniform persistence}

In this section, we study the asymptotic stability of COVID-19-free
equilibrium $V^{0}$ for $R_{c}<1$ and the uniform persistence of system
\eqref{mod1} for $R_{c}>1$.

\begin{theorem}
\label{thm51} The COVID-19-free equilibrium $V^{0}$ is LAS if $\mathcal{R}%
_{c}<1$ and unstable if $\mathcal{R}_{c}>1$.
\end{theorem}

\proof The characteristic equation of the corresponding linearized system of
system \eqref{mod1} at $V^{0}$ can be taken by
\[
F(\Lambda)=(\Lambda+d)^{2}(\Lambda+r_{3}+d)\left(  \Lambda^{3}+b_{1}%
\Lambda^{2}+b_{2}\Lambda+b_{3}\right)  ,
\]
where
\begin{align*}
b_{1}  &  =B_{1}+B_{2}+\left(  c+d\right)  \left[  1-\mathcal{R}_{c}%
+\frac{pc\beta}{(c+d)B_{1}}+\frac{bc\beta(1-p)}{(c+d)B_{2}}\right]  ,\\
b_{2}  &  =\left(  B_{1}+B_{2}\right)  \left(  c+d\right)  \left(
1-\mathcal{R}_{c}\right)  +\frac{B_{2}pc\beta}{B_{1}}+\frac{B_{1}%
bc\beta\left(  1-p\right)  }{B_{2}}+B_{1}B_{2},\\
b_{3}  &  =B_{1}B_{2}\left(  c+d\right)  \left(  1-\mathcal{R}_{c}\right)  ,
\end{align*}
and $\mathcal{R}_{c}$ is calculated as in \eqref{crn}. Obviously,
$F(\Lambda)=0$ has a root $\Lambda=-\left(  r_{3}+d\right)  $ and a double
root $\Lambda=-d$. For $R_{c}<1,$ it is not difficult to find that $b_{1}>0$,
$b_{3}>0$ and $b_{1}b_{2}>b_{3}.$ Therefore, from the Routh-Hurwitz criterion
it follows that any root of the equation $\Lambda^{3}+b_{1}\Lambda^{2}%
+b_{2}\Lambda+b_{3}=0$ has negative real part. That is to say, each root of
$F(\Lambda)=0$ has negative real part, and then $V^{0}$ is LAS.

Obviously, it holds that $b_{3}<0$ for $R_{c}>1$. Hence, there can be found a
positive $\Lambda^{\ast}$ such that $F(\Lambda^{\ast})=0$. In consequence,
$V^{0}$ is unstable. This completes the proof.

From the above discussion, the linear stability of $V^{0}$ follows immediately.

\begin{corollary}
If $\mathcal{R}_{c}=1$, then the COVID-19-free equilibrium $V^{0}$ is linearly stable.
\end{corollary}

The uniform persistence of system \eqref{mod1} has important implications for
controlling the COVID-19 pandemic, which hints that the COVID-19 pandemic will
be persistent with long-term basis. Let $\Omega=\{\phi\in \mathbb{R}_{+}^{6}:\phi_{2}>0\}$ and $u(t)\equiv(S(t),E(t),I(t),A(t),Q(t),R(t))^{T}$ be the
solution of system \eqref{mod1} with any $\phi\in\Omega.$ We can obtain easily
that $\Omega \subseteq D$ is positively invariant for system \eqref{mod1}, and
$u(t)\gg\mathbf{0}$ for $t>0$.

Now, we are in a position to discuss the persistence of system \eqref{mod1} in
$\Omega$. Following the definition in \cite{Butler86,Freedman95}, system
\eqref{mod1} is said to be uniformly persistent if there exists a $\rho>0$
independent of the initial data such that $\rho\leq\liminf_{t\rightarrow
\infty}\psi(t),$ where $\psi=S,E,I,A,Q,R.$ Based on some analysis methods in
\cite{Cheng16,Guo16,Guo18,Guo21,Wang02}, we will give an explicit eventual
lower bound of COVID-19. Now let $\mathcal{R}_{c}>1$, $\eta\in(0,1)$ and
\[
\tilde{S}(\varepsilon)\equiv\dfrac{\lambda}{\eta\beta\left(  aE^{\ast}%
+I^{\ast}+bA^{\ast}\right)  /\left(  S^{0}-\varepsilon\right)  +d}%
,~~\varepsilon\in(0,S^{0}(1-\eta)).
\]
Then there is an $\varepsilon_{0}\in(0,S^{0}(1-\eta))$ such that%
\begin{equation}
\frac{S^{\ast}}{S^{0}}<\frac{\tilde{S}(\varepsilon)}{S^{0}+2\varepsilon}
\label{eq4}%
\end{equation}
for any $\varepsilon\in(0,\varepsilon_{0})$. Note that
\[
\dot{N}(t)=\lambda-dN(t),
\]
we thus have $\lim_{t\rightarrow\infty}N(t)=S^{0}$. Let $S_{\infty}%
=\liminf\nolimits_{t\rightarrow\infty}S(t).$ Then for any $\varepsilon
\in(0,\varepsilon_{0}),$ there exists a $T_{0}\equiv T_{_{0}}\left(
\varepsilon,\phi\right)  >0$ such that for all $t\geq T_{0}$, we have%
\[
I(t)<S^{0}+\varepsilon,\text{ }A(t)<S^{0}+\varepsilon,\text{ }N(t)>S^{0}%
-\varepsilon,\text{ }N(t)<S^{0}+\varepsilon,\text{ }S(t)>S_{\infty
}-\varepsilon.\text{ }%
\]
Let $B=\min\{B_{1},B_{2}\}$ and $m=\max\{a,b,1\}.$ Then $(B+c)/(c+d)$ is
strictly decreasing with respect to $c,$ and it yields that
\begin{equation}
1<\frac{S^{0}}{S^{\ast}}=\bigskip\mathcal{R}_{c}\leq\beta m\frac{B+c}%
{(c+d)B}<\frac{\beta m}{d}. \label{eq5}%
\end{equation}
Hence, for all $t\geq T_{0},$ it follows from the first equation of system
\eqref{mod1} that
\[
\dot{S}(t)>\lambda-\left[  \beta m\left(  1-\frac{S(t)}{N(t)}\right)
+d\right]  S(t)>\lambda-\left[  \beta m\left(  1-\frac{S_{\infty}-\varepsilon
}{S^{0}+\varepsilon}\right)  +d\right]  S(t),
\]
which leads to
\[
S_{\infty}\geq\frac{\lambda}{\beta m\left(  1-S_{\infty}/S^{0}\right)  +d}.
\]
Solving the resulting inequality for $S_{\infty},$ we can obtain $S_{\infty
}\geq\lambda/\beta m$ by means of \eqref{eq5}.

To start the uniform persistence of system \eqref{mod1}, the following lemmas
are needed.

\begin{lemma}
\label{lem1} Assume that $\mathcal{R}_{c}>1$, $\theta\in(0,1)$, and there is a
$t_{0}\geq T_{0}$ such that $E(t)\leq\theta E^{\ast}$ for $t\geq t_{0}$. Then
\[
\frac{S(t)}{N(t)}>\frac{\tilde{k}(\varepsilon)\tilde{S}(\varepsilon)}%
{S^{0}+\varepsilon}>\frac{S^{\ast}}{S^{0}}%
\]
for $t\geq t_{0}+\tilde{T}_{1}(\varepsilon)+\tilde{T}_{2}(\varepsilon),$ where
$\eta\in(\theta,1),$
\begin{align*}
\tilde{T}_{1}(\varepsilon)  &  =\max\left\{  \frac{-1}{B_{1}}\ln\frac
{(\eta-\theta)I^{\ast}}{S^{0}+\varepsilon-\theta I^{\ast}},\frac{-1}{B_{2}}%
\ln\dfrac{(\eta-\theta)A^{\ast}}{S^{0}+\varepsilon-\theta A^{\ast}}\right\}
,\\
\tilde{T}_{2}(\varepsilon)  &  =-\frac{\tilde{S}(\varepsilon)}{\lambda}%
\ln\frac{(1-\tilde{k}(\varepsilon))\tilde{S}(\varepsilon)}{\tilde
{S}(\varepsilon)+\varepsilon-\lambda/\beta m},\text{ }\tilde{k}(\varepsilon
)=\frac{S^{\ast}\left(  S^{0}+2\varepsilon\right)  }{S^{0}\tilde
{S}(\varepsilon)}.
\end{align*}

\end{lemma}

\proof It is not difficult to see that $\tilde{k}(\varepsilon)<1$ from
\eqref{eq4} and $S^{\ast}>\lambda/\beta m$ from \eqref{eq5}. By the third
equation of system \eqref{mod1}, we have
\[
\dot{I}(t)\leq pc\theta E^{\ast}-B_{1}I(t)\text{ for }t\geq t_{0},
\]
which implies that%
\[
I(t)\leq\theta I^{\ast}+\left(  I(t_{0})-\theta I^{\ast}\right)
e^{B_{1}(t_{0}-t)}\leq\theta I^{\ast}+\left(  S^{0}+\varepsilon-\theta
I^{\ast}\right)  e^{B_{1}(t_{0}-t)},
\]
where $I^{\ast}=pcE^{\ast}/B_{1}$. For $t\geq t_{0}+\tilde{T}_{1}%
(\varepsilon),$ it holds $I(t)\leq\eta I^{\ast}.$ Similarly, we have
$A(t)\leq\eta A^{\ast}$ for $t\geq t_{0}+\tilde{T}_{1}(\varepsilon).$ As a
result, it follows that
\[
\frac{aE(t)+I(t)+bA(t)}{N(t)}\leq\frac{\eta\left(  aE^{\ast}+I^{\ast}%
+bA^{\ast}\right)  }{S^{0}-\varepsilon}
\]
for $t\geq t_{0}+\tilde{T}_{1}(\varepsilon),$ and thus there holds%
\begin{align*}
\dot{S}(t)  &  =\lambda-\left[  \dfrac{\beta(aE(t)+I(t)+bA(t))}{N(t)}%
+d\right]  S(t)\\
&  \geq\lambda-\left[  \frac{\eta\beta\left(  aE^{\ast}+I^{\ast}+bA^{\ast
}\right)  }{S^{0}-\varepsilon}+d\right]  S(t)\\
&  =\lambda-\frac{\lambda}{\tilde{S}(\varepsilon)}S(t).
\end{align*}
Consequently, for $t\geq t_{0}+\tilde{T}_{1}(\varepsilon)+\tilde{T}%
_{2}(\varepsilon),$ we have%
\begin{align*}
S(t)  &  \geq\tilde{S}(\varepsilon)+\left(  S(t_{0}+\tilde{T}_{1}%
(\varepsilon))-\tilde{S}(\varepsilon)\right)  e^{-\frac{\lambda}{\tilde
{S}(\varepsilon)}(t-t_{0}-\tilde{T}_{1}(\varepsilon))}\\
&  >\tilde{S}(\varepsilon)+\left(  \frac{\lambda}{\beta m}-\varepsilon
-\tilde{S}(\varepsilon)\right)  e^{-\frac{\lambda}{\tilde{S}(\varepsilon
)}(t-t_{0}-\tilde{T}_{1}(\varepsilon))}\\
&  \geq\tilde{k}(\varepsilon)\tilde{S}(\varepsilon).
\end{align*}
Hence, for $t\geq t_{0}+\tilde{T}_{1}(\varepsilon)+\tilde{T}_{2}%
(\varepsilon),$ it comes to the conclusion that
\[
\frac{S(t)}{N(t)}>\frac{\tilde{k}(\varepsilon)\tilde{S}(\varepsilon)}%
{S^{0}+\varepsilon}>\frac{S^{\ast}}{S^{0}}.
\]

\begin{lemma}
\label{lem2} Under the assumptions of Lemma \ref{lem1}, it holds that
$E(t)\geq\tilde{\nu}=\tilde{\nu}\left(  \varepsilon,t_{0}\right)  \equiv
E(t_{0})e^{\tilde{C}(\varepsilon)\tilde{T}(\varepsilon)}$ for $t\geq t_{0},$
where%
\[
\tilde{T}(\varepsilon)\equiv\max\{\tilde{T}_{1}(\varepsilon)+\tilde{T}%
_{2}(\varepsilon),\tilde{\alpha}(\varepsilon)\},~~\tilde{\alpha}%
(\varepsilon)\equiv\frac{-\ln\left(  1-\tilde{k}(\varepsilon)\right)  }%
{B},\text{ }\tilde{C}(\varepsilon)=\frac{\beta a(\lambda/\beta m-\varepsilon
)}{S_{0}+\varepsilon}-c-d.
\]

\end{lemma}

\proof First, by the second equation of system \eqref{mod1}, we have
\begin{equation}
\dot{E}(t)>\tilde{C}(\varepsilon)E(t).\label{eq6}%
\end{equation}
For $t>t_{0}$, it follows
\begin{equation}
E(t)>E(t_{0})e^{\tilde{C}(\varepsilon)(t-t_{0})}.\label{eq7}%
\end{equation}
Let
\[
\tilde{\nu}=\tilde{\nu}\left(  \varepsilon,t_{0}\right)  \equiv E(t_{0}%
)e^{\tilde{C}(\varepsilon)\tilde{T}(\varepsilon)}.
\]
Then it follows from \eqref{eq7} that $E(t)>\tilde{\nu}$ for $t\in\lbrack
t_{0},t_{0}+\tilde{T}(\varepsilon)]$. For $t>t_{0}+\tilde{T}(\varepsilon
)\text{,}$ we can obtain $E(t)\geq\tilde{\nu}$. In fact, if not, then there is
a $T_{2}>0$ such that $E(t)\geq\tilde{\nu}$ for $t\in\lbrack t_{0},\tilde{t}%
]$, where $\tilde{t}=t_{0}+\tilde{T}(\varepsilon)+T_{2}$, $E(\tilde{t}%
)=\tilde{\nu}$ and $\dot{E}(\tilde{t})\leq0$. Subsequently, we can claim that
$I(\tilde{t})>\tilde{k}(\varepsilon)pc\tilde{\nu}/B_{1}$ and $A(\tilde
{t})>\tilde{k}(\varepsilon)(1-p)c\tilde{\nu}/B_{2}$. Indeed, for $t\in\lbrack
t_{0},\tilde{t}]$, it holds that%
\begin{align*}
\dot{I}(t) &  =pcE(t)-B_{1}I(t)\geq pc\tilde{\nu}-B_{1}I(t),\\
\dot{A}(t) &  =(1-p)cE(t)-B_{2}A(t)\geq(1-p)c\tilde{\nu}-B_{2}A(t).
\end{align*}
And hence,%
\begin{align*}
I(t) &  \geq\dfrac{pc\tilde{\nu}}{B_{1}}+\left(  I(t_{0})-\dfrac{pc\tilde{\nu
}}{B_{1}}\right)  e^{-B_{1}(t-t_{0})}>\dfrac{pc\tilde{\nu}}{B_{1}}\left(
1-e^{-B_{1}(t-t_{0})}\right)  ,\\
A(t) &  \geq\dfrac{(1-p)c\tilde{\nu}}{B_{2}}+\left(  A(t_{0})-\dfrac
{(1-p)c\tilde{\nu}}{B_{2}}\right)  e^{-B_{2}(t-t_{0})}>\dfrac{(1-p)c\tilde
{\nu}}{B_{2}}\left(  1-e^{-B_{2}(t-t_{0})}\right)  .
\end{align*}
Thus, for $t\in\lbrack t_{0}+\tilde{\alpha}(\varepsilon),\tilde{t}],$ we have%
\[
I(t)>\dfrac{\tilde{k}(\varepsilon)pc\tilde{\nu}}{B_{1}},\text{ }%
A(t)>\frac{\tilde{k}(\varepsilon)(1-p)c\tilde{\nu}}{B_{2}}.
\]
The claim is proved.

From Lemma \ref{lem1}, Remark \ref{rem1} and the second equation of system
\eqref{mod1}, it follows
\begin{align*}
\dot{E}(\tilde{t})  &  =\beta\dfrac{S(\tilde{t})}{N(\tilde{t})}(aE(\tilde
{t})+I(\tilde{t})+bA(\tilde{t}))-(c+d)E(\tilde{t})\\
&  >(c+d)\left(  \frac{\tilde{k}(\varepsilon)\tilde{S}(\varepsilon)}%
{S^{0}+\varepsilon}\mathcal{R}_{c}-1\right)  \tilde{\nu}\\
&  >(c+d)\left(  \frac{S^{\ast}}{S^{0}}\mathcal{R}_{c}-1\right)  \tilde{\nu
}=0,
\end{align*}
which contradicts $\dot{E}(\tilde{t})\leq0$. In consequence, $E(t)\geq
\tilde{\nu}$ for $t\geq t_{0}\text{.}$

\begin{lemma}
\label{lem3} Let $\mathcal{R}_{c}>1$ and $\theta\in(0,1)$. Then $\limsup
_{t\rightarrow\infty}E(t)\geq\theta E^{\ast}$.
\end{lemma}

\proof We prove the statement by contradiction. Assume that this is not true.
Then, there exists a $t_{0}\geq T_{0}$ such that $E(t)\leq\theta E^{\ast}$ for
any $t\geq t_{0}$. Now, we define a function as follows,%

\[
L(\phi)=\phi_{2}+\frac{\beta\tilde{k}(\varepsilon)\tilde{S}(\varepsilon
)}{B_{1}\left(  S^{0}+\varepsilon\right)  }\phi_{3}+\frac{b\beta\tilde
{k}(\varepsilon)\tilde{S}(\varepsilon)}{B_{2}\left(  S^{0}+\varepsilon\right)
}\phi_{4},~\phi\in\Omega\text{.}%
\]
Then by Lemma \ref{lem1}, the derivative of $L$ along the solution $u(t)$ for
$t\geq t_{0}+\tilde{T}(\varepsilon)$ can be taken as%
\begin{align*}
\dot{L}(u(t)) ={}  &  \beta\dfrac{S(t)}{N(t)}aE(t)+\frac{\beta\tilde
{k}(\varepsilon)\tilde{S}(\varepsilon)}{B_{1}\left(  S^{0}+\varepsilon\right)
}pcE(t)+\frac{b\beta\tilde{k}(\varepsilon)\tilde{S}(\varepsilon)}{B_{2}\left(
S^{0}+\varepsilon\right)  }(1-p)cE(t)-(c+d)E(t)\\
&  +\beta\left(  \dfrac{S(t)}{N(t)}-\frac{\tilde{k}(\varepsilon)\tilde
{S}(\varepsilon)}{S^{0}+\varepsilon}\right)  I(t)+\beta b\left(  \dfrac
{S(t)}{N(t)}-\frac{\tilde{k}(\varepsilon)\tilde{S}(\varepsilon)}%
{S^{0}+\varepsilon}\right)  A(t)\\
\geq{}  &  (c+d)\left(  \frac{\tilde{k}(\varepsilon)\tilde{S}(\varepsilon
)}{S^{0}+\varepsilon}\mathcal{R}_{c}-1\right)  E(t)\text{.}%
\end{align*}
Consequently, for $t\geq t_{0}+\tilde{T}(\varepsilon)$, it follows from Lemma
\ref{lem2} that
\[
\dot{L}(u(t))\geq(c+d)\left(  \frac{\tilde{k}(\varepsilon)\tilde
{S}(\varepsilon)}{S^{0}+\varepsilon}\mathcal{R}_{c}-1\right)  \tilde{\nu}>0,
\]
which hints $L(u(t))\rightarrow\infty$ as $t\rightarrow\infty\text{.}$
Accordingly, this contradicts the boundedness of $L(u(t))$.

\begin{theorem}
\label{thm1} Suppose $\mathcal{R}_{c}>1$, $\theta\in(0,1)$ and $\eta\in
(\theta,1)$. Then the solution $u(t)$ of system \eqref{mod1} with any $\phi
\in\Omega$ satisfies that
\begin{equation}
\liminf_{t\rightarrow\infty}E(t)\geq\theta E^{\ast}e^{(\frac{ad}{m}%
-c-d)T}=\frac{\theta\lambda\left(  R_{c}-1\right)  }{\left(  c+d\right)
R_{c}}e^{(\frac{ad}{m}-c-d)T}\equiv\nu,\label{einf}%
\end{equation}
where
\begin{align*}
\text{ }T &  =\max\{T_{1}+T_{2},\alpha\},\text{ }\\
T_{1} &  =\max\left\{  \frac{-1}{B_{1}}\ln\frac{\eta/\theta-1}{\left(
1/c+1/d\right)  B_{1}/p\left(  1-1/R_{c}\right)  \theta-1},\frac{-1}{B_{2}}%
\ln\dfrac{\eta/\theta-1}{\left(  1/c+1/d\right)  B_{2}/(1-p)\left(
1-1/R_{c}\right)  \theta-1}\right\}  ,\\
T_{2} &  =-\dfrac{1}{d[\eta(R_{c}-1)+1]}\ln\frac{\left(  1-\eta\right)
(1-1/R_{c})}{1-d\left[  \eta(R_{c}-1)+1\right]  /\beta m},\\
\alpha &  =-\frac{1}{B}\ln[(1-\eta)(1-1/R_{c})].
\end{align*}

\end{theorem}

\proof From Lemma \ref{lem3}, we will consider \eqref{einf} in two cases:
$E(t)\geq\theta E^{\ast}$ or $E(t)$ oscillates around $\theta E^{\ast}$ for
sufficiently large $t$. We thus only need to discuss $E(t)$ oscillates around
$\theta E^{\ast}$. In consequence, we assume that $t_{1}$, $t_{2}\geq T_{0}$
such that
\[
E(t)<\theta E^{\ast}\text{ for }t\in(t_{1},t_{2})\text{ and }E(t_{1}%
)=E(t_{2})=\theta E^{\ast}.
\]
When $t_{2}\leq t_{1}+\tilde{T}(\varepsilon)$, it follows from \eqref{eq6}
that
\[
E(t)>E(t_{1})e^{\tilde{C}(\varepsilon)(t-t_{1})}\geq\theta E^{\ast}%
e^{\tilde{C}(\varepsilon)\tilde{T}(\varepsilon)}=\tilde{\nu}\left(
\varepsilon,t_{1}\right)  =\breve{\nu}>0
\]
for $t\in(t_{1},t_{2}].$ When $t_{2}>t_{1}+\tilde{T}(\varepsilon)$, it holds
$E(t)\geq\breve{\nu}$ for $t\in\lbrack t_{1},t_{1}+\tilde{T}(\varepsilon)]$.
For $t\in\lbrack t_{1}+\tilde{T}(\varepsilon),t_{2}]\text{,}$ then proceeding
exactly as in the proof of Lemma \ref{lem2}, we have $E(t)\geq\breve{\nu}$.
Consequently, $E(t)\geq\breve{\nu}$ for $t\in\lbrack t_{1},t_{2}]\text{.}$
Consider that this kind of interval $[t_{1},t_{2}]$ is chosen arbitrarily.
Thus, $E(t)\geq\breve{\nu}$ for sufficiently large $t\text{,}$ which implies
$\liminf_{t\rightarrow\infty}E(t)\geq\breve{\nu}$. Note that $\varepsilon$ is
given arbitrarily, we thus have $\liminf_{t\rightarrow\infty}E(t)\geq\nu$. In
fact, by Lemma \ref{lem0} and Remark \ref{rem1}, we have%
\[
E^{\ast}=\frac{\lambda\left(  R_{c}-1\right)  }{\left(  c+d\right)  R_{c}%
},\text{ }I^{\ast}=\frac{pc\lambda\left(  R_{c}-1\right)  }{B_{1}\left(
c+d\right)  R_{c}},\text{ }A^{\ast}=\frac{(1-p)c\lambda\left(  R_{c}-1\right)
}{B_{2}\left(  c+d\right)  R_{c}},\text{ }\mathcal{R}_{c}=\frac{S^{0}}%
{S^{\ast}}.
\]
Therefore, it follows
\begin{align*}
\lim_{\varepsilon\rightarrow0^{+}}\tilde{T}_{1}(\varepsilon)= &  \max\left\{
\frac{-1}{B_{1}}\ln\frac{(\eta-\theta)I^{\ast}}{S^{0}-\theta I^{\ast}}%
,\frac{-1}{B_{2}}\ln\dfrac{(\eta-\theta)A^{\ast}}{S^{0}-\theta A^{\ast}%
}\right\}  =T_{1},\text{ }\\
\lim_{\varepsilon\rightarrow0^{+}}\tilde{T}_{2}(\varepsilon)= &  -\frac
{\tilde{S}(0)}{\lambda}\ln\frac{(1-\tilde{k}(0))\tilde{S}(0)}{\tilde
{S}(0)-\lambda/\beta m}=T_{2},\\
\lim_{\varepsilon\rightarrow0^{+}}\tilde{\alpha}(\varepsilon)= &  \frac
{-\ln\left(  1-\tilde{k}(0)\right)  }{B}=\alpha,
\end{align*}
where
\[
\text{ }\tilde{S}(0):=\lim_{\varepsilon\rightarrow0^{+}}\tilde{S}%
(\varepsilon)=\dfrac{S^{0}}{\eta(R_{c}-1)+1},~~\text{ }\tilde{k}%
(0):=\lim_{\varepsilon\rightarrow0^{+}}\tilde{k}(\varepsilon)=\frac{\eta
(R_{c}-1)+1}{R_{c}}.
\]

By Theorem \ref{thm1}, we have the following result immediately.

\begin{theorem}
\label{thm44} If $\mathcal{R}_{c}>1$, then the solution $u(t)$ of system
\eqref{mod1} with any $\phi\in\Omega$ is uniformly persistent, and satisfies%
\begin{align*}
\liminf\limits_{t\rightarrow\infty}S(t)  &  \geq\frac{\lambda}{\beta m},\text{
}\liminf\limits_{t\rightarrow\infty}E(t)\geq\nu,\text{ }\liminf
\limits_{t\rightarrow\infty}I(t)\geq\frac{pc\nu}{B_{1}}=\nu_{1},\text{
}\liminf\limits_{t\rightarrow\infty}A(t)\geq\ \frac{(1-p)c\nu}{B_{2}}=\nu
_{2},\\
\liminf\limits_{t\rightarrow\infty}Q(t)  &  \geq\frac{q_{1}\nu_{1}+q_{2}%
\nu_{2}}{r_{3}+d}=\nu_{3},\text{ }\liminf\limits_{t\rightarrow\infty}%
R(t)\geq\frac{r_{1}\nu_{1}+r_{2}\nu_{2}+r_{3}\nu_{3}}{d}.
\end{align*}

\end{theorem}

In the following, we will exhibit a case to illustrate the distinction of
analysis method of Theorem \ref{thm1}. We reconsider the uniform persistence
for microorganism concentration $m(t)$ of the following microorganism
flocculation model proposed in \cite{Guo18},
\begin{equation}
\left\{
\begin{array}
[c]{l}%
\dot{n}\left(  t\right)  =1-n(t)-\frac{\xi n(t)m(t)}{1+\varrho m(t)},\\
\dot{m}(t)=\frac{\mu n(t-\tau)m(t-\tau)}{1+\varrho m(t-\tau)}-m(t)-\frac
{\gamma m(t)f(t)}{1+\sigma m(t)},\\
\dot{f}(t)=1-f(t)-\frac{\delta m(t)f(t)}{1+\sigma m(t)},
\end{array}
\right.  \label{nmf}%
\end{equation}
where $n(t)$, $m(t)$ and $f(t)$ represent the concentrations of nutrient,
microorganisms and flocculant at time $t$, respectively, the parameters
$\varrho,\sigma\geq0$ and $\tau\geq0$ is time delay, and all other parameters
are positive. Let $u(t)=(n(t),m(t),f(t))^{T}$ be the solution of model
\eqref{nmf} with any $\varphi\in X=\{\varphi\in C([-\tau,0],\mathbb{R}_{+}^{3}%
):\varphi_{2}(0)>0\}$ and the threshold $\mathcal{R}_{0}=\mu/(\gamma+1)>1$ of
model \eqref{nmf}. Then there exists an $\varepsilon_{1}>1$ such that for any
$\varepsilon\in(1,\varepsilon_{1}),$ it follows that%
\[
\tilde{k}(\varepsilon)=\frac{(\varepsilon^{2}\gamma+1)\left[  1+q/(\gamma
+1)\right]  }{(\gamma+1)\mathcal{R}_{0}}<1,\text{ }\liminf
\limits_{t\rightarrow\infty}n(t)\geq\frac{\varrho(\gamma+1)\mathcal{R}_{0}%
+\xi}{(\varrho+\xi)(\gamma+1)\mathcal{R}_{0}}>\frac{\varrho(\gamma
+1)\mathcal{R}_{0}+\xi}{\varepsilon(\varrho+\xi)(\gamma+1)\mathcal{R}_{0}}.
\]

Let $\bar{q}=\bar{q}(\vartheta)=(\gamma+1)\left(  \mathcal{R}_{0}-1\right)
-\vartheta$ for any $\vartheta\in(\bar{\vartheta},(\gamma+1)\left(
\mathcal{R}_{0}-1\right)  )$, where
\begin{equation*}
\bar{\vartheta}=\left\{\aligned
&\max\{(\gamma+1)\left(  \mathcal{R}_{0}-1\right)  -\xi\gamma/\varrho,0\},& \varrho>0,\\
&0,& \varrho=0.
\endaligned
\right.
\end{equation*}
In consequence, we can obtain the following corollary.

\begin{corollary}
\label{th4} If $\mathcal{R}_{0}>1,$ then it holds that
\begin{equation}
\liminf\limits_{t\rightarrow\infty}m(t)\geq\frac{\vartheta}{\left(
\varrho+\xi\right)  (\gamma+1)}e^{-(\gamma+1)\left(  T+\tau\right)  },
\label{yinf}%
\end{equation}
where
\[
T=\frac{1+\varrho\vartheta/\left(  \varrho+\xi\right)  \left(  \gamma+1\right)
}{1+\vartheta/\left(  \gamma+1\right)  }\ln\frac{\gamma/\bar{q}+1/\left[
1+\vartheta/\left(  \gamma+1\right)  \right]  }{\varrho/\xi+1/\left[
1+\vartheta/\left(  \gamma+1\right)  \right]  }.
\]

\end{corollary}

\begin{remark}
In fact, Corollary \ref{th4} is an improvement of \cite[Theorem 4.1]{Guo18}.
By using the method employed in the proof of Theorem \ref{thm1},
the main results on persistence in
\cite{Cheng16,Enatsu12,Guo16,Guo17,Guo21,Guok21,Guok22,Guok220,Li22,Wang14,Zhang09}
can be improved.
\end{remark}

Next, we give a numerical example with Matlab to illustrate the availability
of our work.

\begin{example}
In system \eqref{mod1}, if we take%
\begin{align*}
\lambda &  =1100,\text{ }\beta=0.12,\text{ }a=0.0116,\text{ }b=0.063,\text{
}d=9.6\times10^{-5},\text{ }c=1.2\times10^{-5},\text{ }\\
p &  =0.74,\text{ }q_{1}=0.03,\text{ }q_{2}=0.6,\text{ }r_{1}=0.76,\text{
}r_{2}=0.17,\text{ }r_{3}=0.1.
\end{align*}
Thereby we can obtain $E^{\ast}\approx9.396\times10^{6}$ and $\mathcal{R}%
_{c}\approx12.902>1$ using Matlab. Let the initial data be selected as
$S_{0}=2.1\times10^{7},$ $I_{0}=2.3\times10^{5},$ $E_{0}=4.56\times10^{3},$
$A_{0}=5.76\times10^{3},$ $Q_{0}=1.8\times10^{5},$ $R_{0}=1.8\times10^{5}.$
Then it follows $\liminf\nolimits_{t\rightarrow\infty}E(t)\approx E^{\ast}.$
Take $\theta=0.9$, $\eta=0.901$, we have $\liminf\nolimits_{t\rightarrow\infty
}E(t)>\nu\approx6.725\times10^{6}$ and $\nu/\liminf\nolimits_{t\rightarrow
\infty}E(t)\approx72\%.$ This implies that Theorem \ref{thm1} is valid, and
numerical simulations suggest that the COVID-19 equilibrium $V^{\ast}$ may be
globally attractive if $\mathcal{R}_{c}>1$ in $\Omega$. Among the given
parameters, there can be found $\theta$ and $\eta$ such that $0<\theta<\eta<1$
and $\nu$ is a better estimate of the lower bound on $\liminf
\nolimits_{t\rightarrow\infty}E(t)$. Therefore, $\nu$ is not only a good
explicit estimate of $\liminf\nolimits_{t\rightarrow\infty}E(t)$, but also it
has many practical meanings.
\end{example}

\section{Conclusions}

In this paper, A COVID-19 system \eqref{mod1} with nonlinear incidence rate is
considered. In system \eqref{mod1}, standard incidence rate, instead of
bilinear incidence rate, is used to account for the population growth rate of
exposed individuals, and different quarantined rates and recovery rates for
the symptomatic and asymptomatic infected individuals are introduced. System
\eqref{mod1} admits a unique COVID-19 equilibrium $V^{\ast}$ if and only if
the control reproduction number $\mathcal{R}_{c}>1$. Then the local asymptotic
stability of the COVID-19-free equilibrium $V^{0}$ of system \eqref{mod1} is
proceeded. It shows that if $\mathcal{R}_{c}<1$ (the COVID-19 equilibrium
$V^{\ast}$ is not viable), the COVID-19-free equilibrium $V^{0}$ is LAS, which
implies that COVID-19 pandemic will disappear; if $\mathcal{R}_{c}=1$ (the
COVID-19 equilibrium $V^{\ast}$ is also not viable), the linearized system of
system \eqref{mod1} at $V^{0}$ is stable; if $\mathcal{R}_{c}>1$, $V^{0}$ is unstable.

For persistence dynamics of system \eqref{mod1}, it shows that $V^{\ast}$ is
viable, i.e., $\mathcal{R}_{c}>1$, the COVID-19 pandemic is uniformly
persistent. Furthermore, a more refined analysis method is proposed to better
estimate the ultimate lower bound of COVID-19 infected individuals if
$\mathcal{R}_{c}>1,$ which also non-trivially improves some analysis
techniques for system persistence in
\cite{Cheng16,Enatsu12,Guo16,Guo17,Guo18,Guo21,Guok21,Guok22,Guok220,Li22,Wang02,Wang14,Zhang09} as well as can be
applied to other related mathematical models in biology.
It is not difficult to find that $\mathcal{R}_{c}$ is a decreasing function
with respect to the quarantined rates $q_{1}$ and $q_{2}.$ We thus can
strengthen the quarantine, which also effectively reduce COVID-19
transmission. In addition, numerical simulations show that the COVID-19-free
equilibrium $V^{0}$ and the COVID-19 equilibrium $V^{\ast}$ may be globally
attractive for $\mathcal{R}_{c}<1$ in $D$ and $\mathcal{R}%
_{c}>1$ in $\Omega,$ respectively. Therefore, the global stability problems of
$V^{0}$ and $V^{\ast}$ are very practical and challenging, which means that
the COVID-19 pandemic will die out or be persistent under certain conditions,
we will settle these problems in future work.

\section*{Acknowledgements}

This work is partially supported by the National NSF of China (Nos. 11901027,
11871093, 11671382, 11971273 and 12126426), the Major Program of the National
NSF of China (No. 12090014), the State Key Program of the National NSF of China
(No. 12031020), the NSF of Shandong Province (No. ZR2018MA004), and the China
Postdoctoral Science Foundation (No. 2021M703426),
and the Pyramid Talent Training Project of BUCEA (No. JDYC20200327).

\section*{Conflict of interest and data availability statements}

The authors declare that they have no conflict of interest, and data sharing
is not applicable to this article as no datasets were generated or analysed
during the current study.

\section*{References}

%\bibliography{mybibfile}

\end{document}